\pdfoutput=1

\documentclass[pdftex,moreauthors]{Definitions/mdpi} 
\preto{\abstractkeywords}{\nolinenumbers}
\nolinenumbers
\usepackage[utf8]{inputenc}
\usepackage[T1]{fontenc}
\usepackage{lmodern}
\usepackage{geometry}
\usepackage{amsmath,amssymb}
\usepackage{graphicx}
\usepackage{booktabs}
\usepackage{natbib}
\usepackage{authblk}
\usepackage{luatex85}

\def\apjl{ApJL}
\def\apj{ApJ}
\def\pasj{PASJ}
\def\aap{A\&A}
\firstpage{1} 
\pubvolume{1}
\issuenum{1}
\articlenumber{0}
\pubyear{2026}
\copyrightyear{2026}
\datereceived{ } 
\daterevised{ } 
\dateaccepted{ } 
\datepublished{ } 
\Title{Flagging Super-Eddington Candidates among Jetted, \texorpdfstring{$\gamma$}{gamma}-Ray-Emitting AGN}

\Author{Paola Marziani $^{1,2,\dagger}$\orcidA{}, Benedetta Dalla Barba$^{3}$\orcidB{}, Luigi Foschini$^{3}$\orcidC{}  
}

\address{%
$^{1}$ \quad {Astronomical Observatory of Padova, National Institute for Astrophysics (INAF), }
  35122 Padova, Italy \\
  $^{2}$ \quad Instituto de Astrof\'\i sica de Andaluc\'\i a (IAA), Consejo Superior de Investigaciones Cient\'{\i}ficas {(CSIC)}
, Glorieta de Astronom\'\i a s/n, ES18008 Granada, Spain\\
$^3$ \quad Osservatorio Astronomico di Brera, Istituto Nazionale di Astrofisica (INAF), Via E. Bianchi 46, Merate (LC) 23807, Italy    }

\firstnote{Presented at the 2026 Electronic Conference on Universe (IOCU 2026), 4–6 March 2026; Available online: \url{https://sciforum.net/event/IOCU2026}.}

\abstract{
The quasar Eigenvector-1/Main Sequence (E1/MS) provides a physically motivated empirical framework to organize the spectroscopic diversity of type~1 active galactic nuclei (AGN). In its optical plane, the full width at half maximum of H$\beta$ and the Fe\,II strength ratio $R_{\mathrm{FeII}}$ define a sequence that is primarily driven by Eddington ratio, with important secondary roles played by black-hole mass, orientation, spectral energy distribution, and chemical enrichment. The E1/MS framework is therefore well suited to identify highly accreting and possibly super-Eddington (SE) sources, usually associated with the extreme Population~A (xA) spectral types. We discuss why E1/MS is a useful tool to search for SE accretors among jetted AGN and, conversely, to place $\gamma$-ray-detected AGN in the broader context of quasar phenomenology. We summarize two complementary results: (1) some candidate SE accretors show radio properties {  such as high brightness temperature non-thermal cores or radio lobes} consistent with jet activity; and (2) a subset of low-redshift $\gamma$-ray narrow-line Seyfert~1 galaxies exhibit optical spectra consistent with xA or borderline-xA classification. We also expand the discussion of recent developments in E1/MS studies, including metallicity trends, the spectral energy distribution of xA quasars, and the role of highly accreting quasars   {  as discovery tools for extreme accretion states, as probes of quasars at the reionization epoch, and as possible cosmological probes.}}

\begin{document}

\section{Introduction}

The diversity of type~1 AGN is not random. Over the past three decades, it has become increasingly clear that a large fraction of the observed phenomenology can be organized along a low-dimensional parameter space, originally revealed by principal component analysis and now commonly referred to as Eigenvector~1 (E1) or the quasar Main Sequence (MS) \citep{BorosonGreen1992,sulenticetal00a,sulenticetal00c,MarzianiSulentic2014,Panda2024}. The practical importance of E1/MS is that it connects easily measured spectral quantities to the physical state of the accreting black hole system.

The optical plane of the sequence is defined by the full width at half maximum (FWHM) of broad H$\beta$ and the intensity ratio
$
R_{\mathrm{FeII}}=\frac{I(\mathrm{FeII}\ \lambda 4570)}{I(\mathrm{H}\beta)},
$  where the Fe\,II blend is measured over the 4434--4684~\AA\ interval. Population~A sources occupy the narrower-line side of the sequence, while Population~B sources populate the broader-line side, with the transition conventionally placed at $\mathrm{FWHM}(\mathrm{H}\beta)\approx 4000~\mathrm{km\,s^{-1}}$ for low-luminosity quasars \citep{Sulenticetal2000,marzianietal18}. The extreme Population~A (xA) sources, usually selected by $R_{\mathrm{FeII}}\gtrsim 1$, mark the high-accretion end of the sequence and are widely regarded as the best observational candidates for near- or super-Eddington accretion \citep{duetal16a,pandamarziani23,marzianietal25}.

The present contribution emphasizes a specific question: can SE accretion coexist with a relativistic jet? The question is timely because  powerful $\gamma$-ray emission is widely interpreted as a signature of a radio-loud AGN, and therefore of a relativistically jetted central engine\footnote{Hereafter, “relativistically jetted” (or simply “jetted”) will be preferentially used in place of “radio-loud”.}, while xA sources are traditionally associated with strong radiation fields, dense line-emitting gas, and powerful winds rather than with classical jetted phenomenology.  Models of super-Eddington accretion around supermassive black holes are capable of producing only mildly relativistic outflows \citep{ohsugaetal05,takeuchietal13,ogawaetal17,kitakietal18,jiangetal19}. The results summarized here suggest that the overlap between these regimes is real, even if uncommon. In the following sections, we preliminarily  expand the discussion of the E1/MS context in light of recent developments.

\section{The E1/Main Sequence framework and recent developments}
\label{sec:e1}

\subsection{Physical meaning of the optical plane}

In its simplest representation, the quasar MS is not merely a classification diagram, but a projection of several coupled physical parameters \citep{MarzianiSulentic2014,Panda2024}. The dominant driver is generally identified with the Eddington ratio, while black-hole mass, viewing angle, and line-emitting gas properties modulate the observed location of an individual source \citep{marzianietal01,marzianietal18}. Along the sequence from Population~B to xA, one usually observes a strengthening of optical Fe\,II emission, a weakening of narrow [O\,III], and increasingly frequent or stronger blueshifts in high-ionization lines, all of which are consistent with a growing role of radiative driving and outflows \citep{MarzianiSulentic2014,pandamarziani23}.

The xA sources are especially relevant because they appear to isolate a regime in which the accretion flow is radiatively efficient, geometrically thick in the innermost region, and accompanied by strong winds. Their spectra are often recognizable by eye: strong Fe\,II, weak [O\,III], Lorentzian or nearly Lorentzian broad Balmer profiles, and large C\,IV blueshifts when ultraviolet data are available \citep{duetal16a,marzianietal25}. These empirical properties make xA quasars practical targets for identifying SE candidates in large optical surveys {  \citep{Negreteetal2018,matthewsetal19,lambridesetal26}.}

\subsection{Chemical enrichment and the role of metallicity}

One of the most important recent developments is the increasing evidence that chemical abundance is not a minor detail but a significant correlate of the MS. A detailed analysis of optical and ultraviolet line diagnostics has shown that metallicity increases systematically along the sequence, reaching several times solar and in some cases tens of solar in high-Eddington sources \citep{Florisetal2024}. This result has two major implications. First, it supports the use of Fe\,II prominence as a physically meaningful tracer, rather than a purely phenomenological quantity. Second, it strengthens the view that the xA regime may be associated with intense circumnuclear star formation and rapid chemical processing, which in turn can influence the ionizing continuum, cooling balance, and line-driving efficiency \citep{Florisetal2024,marzianietal25}.

The metallicity trend also helps explain why xA sources often show extreme ultraviolet line ratios and strong winds. Higher metal abundance enhances radiative coupling and can facilitate wind launching, while at the same time affecting the cooling pathways of the line-emitting gas {  \citep{murrayetal95,murraychiang97,proga07}. } In this sense, the E1/MS sequence increasingly appears to encode a coupled evolution of accretion state, gas physical conditions, and chemical enrichment {  \citep{fraix-burnetetal17,pandaetal19,marzianietal25,naddafetal25}. Multi-frequency observational data support the presence of different accretion modes in Populations A and B: a geometrically thin, optically thick disk for Pop. B  \citep{shakurasunyaev73}, while for Pop. A and especially extreme Pop. A the data are consistent with the appearance of an optically thick advection-dominated accretion flow \citep{abramowiczetal88,abramowicz05}. The presence  of a geometrically thick structure provides a shielding  effect that makes  efficient wind acceleration possible \citep{giustiniproga19}, the low-ionization state and the remarkable optical stability of the virialized region \citep{duetal14,duetal15}. }

\subsection{The spectral energy distribution of xA quasars}

A second major advance is the derivation of an empirical spectral energy distribution (SED) tailored specifically to xA quasars. \citet{Garnicaetal2025} constructed an SED for extreme Population~A sources, showing that their broadband shape differs in meaningful ways from average quasar templates: the big blue bump is more prominent, the optical/UV continuum is stronger, and the X-ray continuum tends to be steep. These results are important because photoionization calculations for xA sources should not rely blindly on generic quasar continua.

The new xA SED helps connect the E1 classification to the microphysics of line formation. In practical terms, it offers a more appropriate continuum for computing ionization, radiation pressure, and line emissivities in high-accretion quasars. This is especially relevant for assessing whether strong Fe\,II, weak [O\,III], and prominent blueshifts can be reproduced self-consistently within a single physical scenario. The SED results also reinforce the idea that xA quasars form a spectroscopically coherent population, rather than a loose collection of outliers \citep{Garnicaetal2025}.

\subsection{E1/MS as a tool beyond classification}

The E1/MS framework has also matured into a discovery and interpretation tool. High-Eddington quasars selected through E1/MS criteria have been proposed as laboratories for extreme accretion physics, for broad-line region (BLR) structure studies, and even for cosmological applications \citep{pandamarziani23,marzianietal25}. The logic is simple: once one isolates a population with small dispersion in accretion state and line properties, the population becomes suitable for testing scaling relations or for constructing luminosity estimators with reduced intrinsic scatter \citep{Negreteetal2018}.

This broader use of the MS is especially relevant to the present work. If xA sources can be identified robustly from optical spectra, then the same framework can be applied to sources selected by radio or $\gamma$-ray properties, providing an immediate way to ask whether a jetted object also belongs to the highest-accretion part of the sequence.

\section{Why use \texorpdfstring{$\gamma$}{gamma}-ray emission and radio loudness?}

Powerful $\gamma$-ray emission in AGN is most naturally explained by inverse-Compton processes within a relativistic jet. In practice, a confirmed $\gamma$-ray detection identifies a jetted source and therefore a radio-loud or, more precisely, a jet-dominated system \citep{foschini26}. This is a powerful selection because it bypasses some of the ambiguities of radio loudness based solely on radio-to-optical flux ratios, which can be affected by host-galaxy contamination, beaming, and variability.

Within this framework, the comparison explored here becomes especially meaningful. If E1/MS identifies the highest-accretion sources, and $\gamma$ rays identify relativistic jets, then the overlap between the two selections isolates rare systems in which extreme accretion and jet production coexist. Conversely, the absence of overlap would support the view that strong radiation pressure and large BLR covering factors suppress or inhibit jet formation. The current observational picture suggests that the overlap exists, but it is small and likely heterogeneous.

\begin{figure}[htbp]
\begin{center}
\includegraphics[scale=0.25]{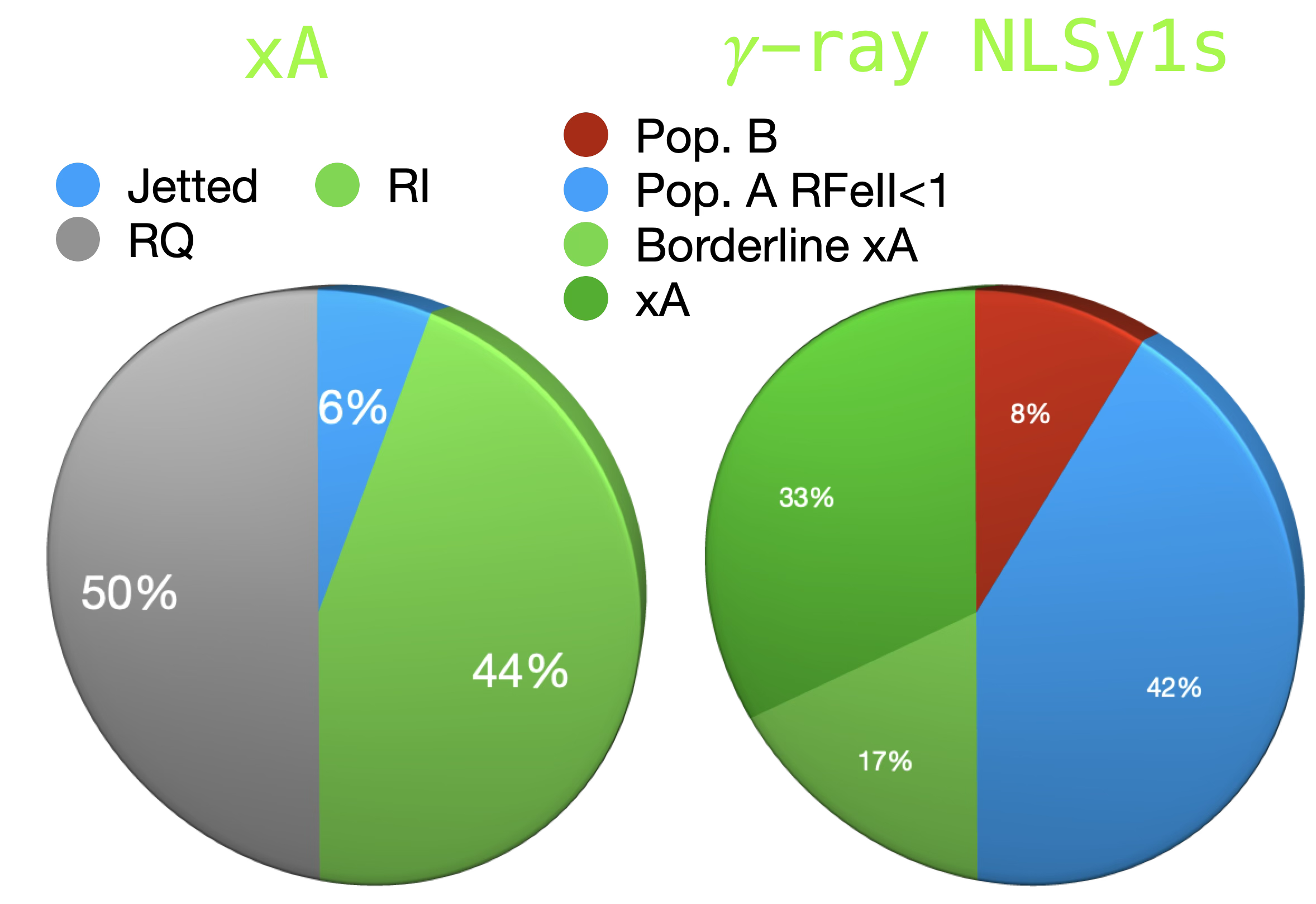}
\caption{{  Left: prevalence of jetted (RL, with ratio between radio emission at 6 cm and $g$-band specific flux $R_\mathrm{K} \gtrsim 60$, radio-intermediate (RI, $10 \lesssim R_\mathrm{K} \lesssim 60$) and radio-quiet (RQ, $R_\mathrm{K} \lesssim 10$). Right: prevalence of optical spectral type among $\gamma$-ray-emitting NLSy1s.} $\gamma$-ray detection effectively identifies a jetted source and thus confirms its RL nature.   One jetted source (following the strict criterion of \citet{zamfiretal08}) was found in the xA sample of \citet{gendron-marsolaisetal26}.  At least four sources with highly distinguishable xA spectra were found in a sample of low-z,  $\gamma$-ray emitting NLSy1s by \citet{DallaBarbaetal2026}.  }
\label{fig:pies}
\end{center}
\end{figure}

\section{Two complementary observational approaches}

\subsection{Jetted sources among super-Eddington candidates}

The first approach starts from candidate SE accretors and investigates whether their radio properties require a jetted interpretation. Recent work on 18 SE candidates with radio emission has shown that the radio output cannot always be ascribed to a single mechanism: in many cases star formation, {  unresolved  AGN activity due to coronal emission}, and jet-related emission may all contribute \citep{gendron-marsolaisetal26}. However, the important point for the present discussion is that at least one source in that sample satisfies a strict jetted criterion while simultaneously belonging to the xA domain in the optical MS, following the conservative radio-loudness separation discussed by \citet{zamfiretal08}.

This result is important even if the number of clear cases is still small. It demonstrates that the coexistence of a relativistic jet with xA-like, presumably SE accretion is not forbidden observationally. The rarity of such sources may then reflect either a short duty cycle, strong selection effects, or a genuine tension between sustained jet activity and the conditions typical of the xA BLR \citep{czernyetal09,foschini25}.

\subsection{xA-like spectra among \texorpdfstring{$\gamma$}{gamma}-ray-emitting NLSy1 galaxies}

The second approach starts from jetted AGN, specifically $\gamma$-ray-emitting narrow-line Seyfert~1 (NLSy1) galaxies, and asks whether some of them occupy the xA domain in E1/MS. Optical follow-up of $\gamma$-ray-selected AGN has recently increased the number of optically confirmed $\gamma$-ray NLSy1s and refined their classifications \citep{DallaBarbaetal2026,foschini26}. Within this context,  at least four low-redshift $\gamma$-ray NLSy1s show distinctive xA spectra, or spectra that are very close to the xA border.

This is again a non-trivial result. NLSy1 galaxies are often associated with relatively low black-hole masses and high Eddington ratios, but not all NLSy1s are xA sources, and not all jetted NLSy1s display the strong Fe\,II, weak [O\,III], and wind-dominated signatures typical of xA. The identification of a subset of $\gamma$-ray NLSy1s with clear xA-like spectra therefore indicates that some jetted Seyfert nuclei do reach the most extreme end of the MS, despite the expectation that jet launching, mass loading, and radiative drag might complicate or suppress such a regime.

\section{Discussion}

\subsection{Can super-Eddington accretion and relativistic jets coexist?}

The empirical answer appears to be yes. The overlap between the two selections is not large, but it is now difficult to argue that the two regimes are mutually exclusive. From the E1/MS perspective, the significance of the overlap is twofold. First, it shows that the xA classification remains meaningful even in the presence of strong non-thermal activity, provided that the optical spectrum is of sufficient quality and the continuum and Fe\,II contributions are modelled appropriately. Second, it suggests that the physical conditions usually associated with xA sources --- strong radiation fields, dense BLR gas, significant line driving, and possibly thick inner flows --- do not necessarily prevent the formation or maintenance of a relativistic jet.

At the same time, the scarcity of clear cases hints that coexistence may  require special circumstances.  AGN that are both jetted and super-Eddington accreting may represent the intersection of two minority properties, $\sim 10$\% of type 1 AGN being radio-loud (RL), and $\sim 10$ \% being super-Eddington candidates, with an expected prevalence of $\approx 1$\%, if  the two properties are fully independent. However, the jet may be intermittent, or its observed prominence may depend strongly on orientation and Doppler boosting.  Conversely, the xA classification may itself be easier to recognize when the thermal continuum and broad-line spectrum are not overwhelmed by a strongly beamed synchrotron component. A full census will therefore require coordinated optical, radio, and high-energy studies.

\subsection{Implications for BLR and jet physics}

The overlap between xA and jetted sources is interesting because the xA regime likely maximizes the density of seed photons from the accretion flow, BLR, and torus. In a jetted system, such photon fields can boost external-Compton emission and thus favor $\gamma$-ray detectability. This possibility is attractive because it naturally links the high radiative efficiency of xA sources to their high-energy phenomenology. On the other hand, dense line-emitting gas and strong radiation pressure may increase mass loading or radiative drag, making it more difficult for a powerful jet to remain clean and highly relativistic. The observational coexistence of both ingredients therefore provides a useful benchmark for numerical models of disk--wind--jet coupling \citep{marzianietal25,gendron-marsolaisetal26}.

\subsection{Future prospects}

Several developments make this line of work timely. Large optical spectroscopic surveys can identify xA candidates efficiently, while Fermi-LAT and radio interferometric surveys continue to improve the census of jetted Seyfert-like AGN \citep{foschini26}. At the same time, the refinement of xA SEDs and metallicity estimates opens the way to more realistic photoionization and radiative-transfer modelling \citep{Florisetal2024,Garnicaetal2025}. The next step is to move from qualitative overlap to quantitative population studies: what fraction of jetted Seyfert nuclei are xA? What fraction of xA sources host compact or large-scale jets? How do black-hole mass, orientation, metallicity, and host-galaxy properties regulate the overlap?

\section{Conclusions}

The E1/Main Sequence remains the most practical empirical roadmap for identifying the most highly accreting type~1 AGN. In the optical plane, xA quasars occupy the high-$R_{\mathrm{FeII}}$ end of the sequence and provide the most direct route to flagging SE candidates. Recent work has strengthened this interpretation by showing that metallicity increases systematically toward the xA extreme, and by deriving a broadband SED specifically tailored to these sources.

This proceedings contribution highlights two complementary results. First, some candidate SE accretors with radio emission show evidence for genuine jet activity. Second, a subset of low-redshift $\gamma$-ray NLSy1 galaxies display xA or borderline-xA optical spectra. Taken together, these findings indicate that SE accretion and relativistic jets can coexist, even if the combination is uncommon and probably subject to strong selection effects.

The broader implication is that E1/MS provides a common language for linking optical spectroscopy, radio loudness, and high-energy emission. This makes it especially valuable for identifying rare systems in which accretion, winds, and jets all operate close to their most extreme observed limits.

\section*{Funding}
PM acknowledges financial support from the Spanish MCIU through project PID2022-140871NB-C21 by “ERDF A way of making Europe”, and from the Severo Ochoa grant CEX2021-515001131-S funded by MCIN/AEI/10.13039/501100011033.

\section*{Institutional Review Board Statement}
Not applicable.

\section*{Informed Consent Statement}
Not applicable.

\section*{Data Availability Statement}
No new data were created for this proceedings contribution. The manuscript is based on results presented in the cited references. 

\section*{Conflicts of Interest}
The author declares no conflict of interests.

\bibliographystyle{plainnat}

\end{document}